# Simulation of water flow and solute transport as affected by a soil layer interface and a subsurface drainage system using a numerical model

Amir Sedaghatdoost[1]

[1]PhD student, Department of Biological and Agricultural Engineering, Texas A&M University, College Station, TX, 77843, USA.

**Abstract**

Vadose zone consists of various heterogeneities and boundary conditions, which profoundly impact the water flow and solute transports in the soil. Understanding these effects is vital for predicting biogeochemistry of the soil and preventing environmental issues associated with groundwater pollution. The HP1 model has been applied for numerical analysis of various water flow and chemical transport problems in the soil. The objective of the paper was to assess the effects of a soil layer and a subsurface drainage system on the water flow and solute transport in a soil column experiment using the HP1 model. Results indicated that both the soil layer and the subsurface drainage system had profound impacts on the geochemical processes in the soil. The soil layer retarded the transport of solute in the soil and created a hotspot for geochemical processes. On the other hand, the subsurface drainage system drained out an excessive amount of water and solutes from the soil and therefore decreased the solute concentration in deep soil.

# 1 Introduction

Soils are complex three-dimensional heterogeneous systems involving the transport, redistribution, and transformation of multiple components in different phases under non-isothermal conditions. The various transformations are, to a large extent, initiated, directed, and catalyzed by physical, chemical, and biological processes. Although data-driven methods recently extended, presented, and applied by many research scholars in the area of geotechnical engineering (Moayedi et al. 2018) as well as prediction of water flow (Araghinejad et al. 2018), they are hardly able to describe the complex behavior of soil. An integrated approach is necessary for understanding and unraveling the complex interplay of different soil processes. Coupling physical and geochemical processes within one integrated numerical simulator provide a process-based tool for investigating the mobility of contaminants as affected by changing hydrologic regimes and geochemical conditions.

Numerous models can be used to simulate the transport of chemical substances into the vadose zone (Mokari et al., 2019). One such model that has been successfully applied in many different geochemical problems is HP1. This one-dimensional model has been built by the coupling PHREEQC geochemical program with HYDRUS-1D. HP1 is a holistic model that can be applied to solve problems related to variably saturated flow conditions, the fate and transport of multiple components, heat transport, and various biogeochemical reactions. The major drawback of HP1 is the number of input parameters required by the model. This issue can be overcome by methods such as inverse modeling approach (Sedaghatdoost et l., 2018; Sedaghtdoost et al., 2019), which is estimating input parameters using a few measured data and outputs of the model (Sedaghatdoost and Ebrahimian, 2015).

Liu et al. (2013) applied HP1 to assess the impact of long-term irrigation practices on sulfate transport to the groundwater when different sources of irrigation water were used. They showed that simulated water contents closely mirrored measured values at all depths. Schonsky et al. (2013) were also showed that HP1 could accurately simulate concentrations of sulfate and bromide during a percolation experiment test. Zhang et al. (2013) applied a semi-reactive microbial transport model within HP1 and used it to a column transport experiments with constant and variable solution chemistries. They showed that the model matched observed bacterial breakthrough curves well. Moreover, although limitations exist in the application of a semi-reactive microbial transport model, this method represents one step towards a more realistic model of bacterial transport in complex microbial–water–soil systems. Leterme and Jacques (2015) applied HP1 for the transport of Hg in sandy soil under different scenarios.

The goal of this paper is to realize the effect of different domain properties on the simulation of water and multi-component solute transport in a variably saturated media. In particular, the effect of soil layers and a subsurface drainage system on the results of HP1 will be evaluated. Moreover, the aim of this project is to see if soil heterogeneities can be considered as hot spots for biogeochemical processes. There have been no studies that applied horizontal drains as the lower boundary condition in HP1. Therefore, this is a new idea to use HP1 to evaluate the leaching of solutes through a subsurface drainage system.

# 2 Material and Methods

## 2.1 Model description

HP1 uses the Richards equation and Mualem-van Genuchten model to simulate water flow and define soil hydraulic properties, respectively. The solute transport in the model is governed by aqueous master species, which significantly decrease the computational cost of the model. A summary of governing flow and transport equations in the HP1 model is provided in table 1.

## 2.2 Experimental setup

- *Homogeneous soil*

The experiment was conducted in a 50-cm soil column, which had a dry loamy soil with an initial pressure head of -300 cm. Infiltration occurred under a constant upper boundary pressure head of –50 cm. There was a free drainage condition for the bottom boundary condition. The following elements were considered: Br, Ca, Cd, Cl, K, Mg, and Na. The bulk density was 1.31 g/cm3, and CEC is 0.05371 mol / 1000 $cm^3$ soil. Initial concentrations were: [Cl] = 69 μmol/kg water, [Ca] = 6 μmol/kg water, [K] = 4 μmol/kg water, [Na] = 64 μmol/kg water, [Mg] = 8 μmol/kg water, [Cd] = 0.8 μmol/kg water and [Br] = 62 μmol/kg water. The inflowing solution had following composition: [Ca] = 0.01 mol/kg water and [Cl] = 0.02 mol/kg water.

- *Two-layered soil*

In this experiment, the soil was assumed to have two layers of 25 cm loamy sand on a 25 cm silty clay loam. Five observation nodes were used to evaluate the effect of soil layer on the hydrology and chemistry of soil profile.

- *Two-layered soil with a tile drainage system*

In this experiment, subsurface drainage was used instead of free drainage conditions in previous experiments. The drainage system consisted of a 25 cm deep tile drainage system that had 1000 cm spacing between lateral drains. Five observation nodes were used to assed the impact of the subsurface drainage system on the solute concentrations and hydrology of the experiment.

# 3 Results and Discussions

## 3.1 Two-layered soil

Figure 1 showed that the presence of a soil layer resulted in a significant difference between the results of observation nodes above and below the soil layer interface. It demonstrated that nodes below the soil layer had higher water content as they had a finer soil. On the other hand, the water fluxes from upper nodes were higher since they had a coarser soil and thus could transfer water faster. Moreover, figure 2 demonstrated that the soil layer had a profound effect on the water flow in the soil profile. It indicated that the soil below the layer interface had higher water content, in comparison with upper soil, during the study period. As can be seen, there was a sharp change in the amount of water content, and pressure heads, which resulted from the sharp difference between two different soil types existed above and below the soil layer interface.

**Table 1. Governing flow and transport equations, equilibrium geochemical mass-action equations, and examples of kinetic reaction equations used in HP1.**

| Process | Equation† | Equation no. |
|---|---|---|
| | Flow and Transport Equations | |
| Water flow (Richards equation) | $\frac{\partial \theta(h)}{\partial t} = \frac{\partial}{\partial x}\left[K(h)\left(\frac{\partial h}{\partial x} + \cos\alpha\right)\right] - S(h)$ | [1.1] |
| Heat transport | $\frac{\partial C_p(\theta)T}{\partial t} = \frac{\partial}{\partial x}\left[\lambda(\theta)\frac{\partial T}{\partial x}\right] - C_w\frac{\partial qT}{\partial x} - C_w ST$ | [1.2] |
| Solute transport (advection-dispersion equation) | $\frac{\partial \theta C_j}{\partial t} = \frac{\partial}{\partial x}\left(\theta D^w \frac{\partial C_j}{\partial x}\right) - \frac{\partial q C_j}{\partial x} - SC_{r,j} + R_{o,j}$ | [1.3] |
| | Chemical Equilibrium Reactions | |
| | Reaction Equation; Mass Action Law | |
| Aqueous speciation | $\sum_{j=1}^{N_m} \nu_{ji}^{l} A_j^{m} = A_i$; $K_i^{l} = \gamma_i^{l} c_i \prod_{j=1}^{N_m}\left(\gamma_j^{m} c_j^{m}\right)^{-\nu_{ji}^{l}}$ | [1.4] |
| Ion exchange | $\sum_{j=1}^{N_m} \nu_{ji_e}^{e} A_j^{m} + \nu_{j_e i_e}^{e} X_{j_e}^{m} = X_{i_e}$; $K_{i_e}^{e} = \gamma_{i_e}^{e} \beta_{i_e,f_e}^{e} \prod_{j=1}^{N_m}\left(\gamma_j^{m} c_j^{m}\right)^{-\nu_{ji_e}^{e}} \left(\gamma_{j_e}^{e} \beta_{j_e,f_e}^{e}\right)^{-\nu_{j_e i_e}^{e}}$ | [1.5] |
| Surface complexation | $\sum_{j=1}^{N_m} \nu_{ji_s}^{s} A_j^{m} + \nu_{j_s i_s}^{s} S_{j_s}^{m} = S_{i_s}$; $K_{i_e}^{s,int} = \left[\beta_{i_s,f_s}^{s} \prod_{j=1}^{N_m}\left(\gamma_j^{m} c_j^{m}\right)^{-\nu_{ji_s}^{e}} \left(\beta_{j_s,f_s}^{s}\right)^{-\nu_{j_s i_s}^{s}}\right] \exp\left(\frac{F\Psi_{f_s}}{RT}\Delta z_{i_s}\right)$ | [1.6] |
| Mineral dissolution | $\sum_{j=1}^{N_m} \nu_{ji_p}^{p} A_j^{m} = M_{i_p}$; $K_{i_p}^{p} = \prod_{j=1}^{N_m}\left(\gamma_j^{m} c_j^{m}\right)^{-\nu_{ji}^{p}}$ | [1.7] |
| | Chemical Kinetic Reactions | |
| General rate equation for aqueous species | $\frac{dc_i}{dt} = \sum_{j_k=1}^{N_k} \nu_{j_k,i}^{k} R_{j_k}$ | [1.8] |
| Mineral dissolution equation (Lasaga, 1995) | $R_{j_k} = k_0 A_{min} \exp\left(-E_a/RT\right) a_{H^+}^{n_{H^+}} g(I) \prod_i a_i^{n_i} f(\Delta G_r)$<br>affinity term based on transition state theory (Aagaard and Helgeson, 1982; Lasaga, 1981, 1995, 1998):<br>$f(\Delta G_r) = 1 - Q/K_{i_p}^{p}$ | [1.9] |
| Monod rate equation (Schäfer et al., 1998) | $R_{j_k} = v_{max,j_k} X_{mo,r} \prod_{m=1}^{N_{M,jk}} \frac{c_{j_k,m}}{K_{j_k,m}^{M} + c_{j_k,m}} \prod_{i=1}^{N_{I,jk}} \frac{K_{j_k,i}^{I}}{K_{j_k,i}^{I} + c_{j_k,m}}$ | [1.10] |

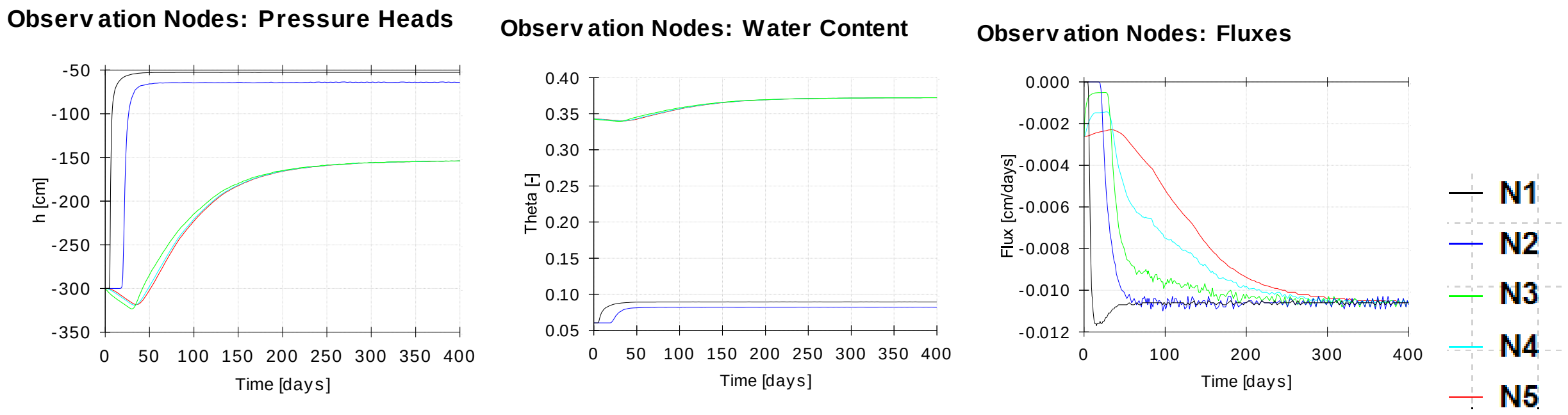

**Fig. 1 variation of pressure heads and water content in different observational nodes of the two-layered soil profile**

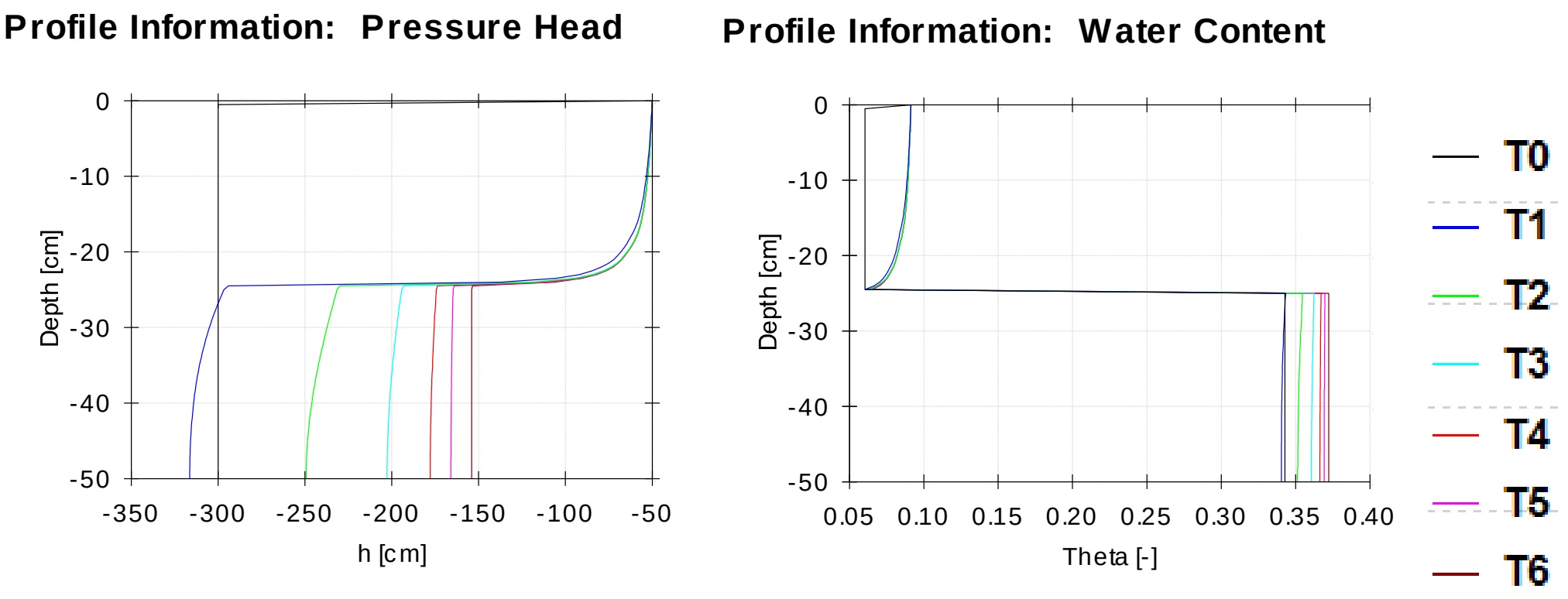

**Fig. 2 pressure head and water content at different times of the experiment in the two-layered soil profile**

Results showed that the soil layer makes cations accumulate in the layer interface before they moved to deeper soil (figure 3). Figure 6 showed that the concentration of most cations decreases by time and depth; however, there was a nearly constant concentration between 5 cm and 25 cm, which may occur because of the presence of the soil layer. This area, as Hansen et al. (2011) mentioned, acted as a hotspot for geochemical processes that highly impact the transport of solutes in the soil. In contrast to cations, anions transport was not affected by the soil layer considerably. This might arise from the fact that anions do no usually adsorb to soil particles and do not tend to make any chemical reaction with clay particles. Accordingly, the change in the soil type may not affect their transport as much as cations.

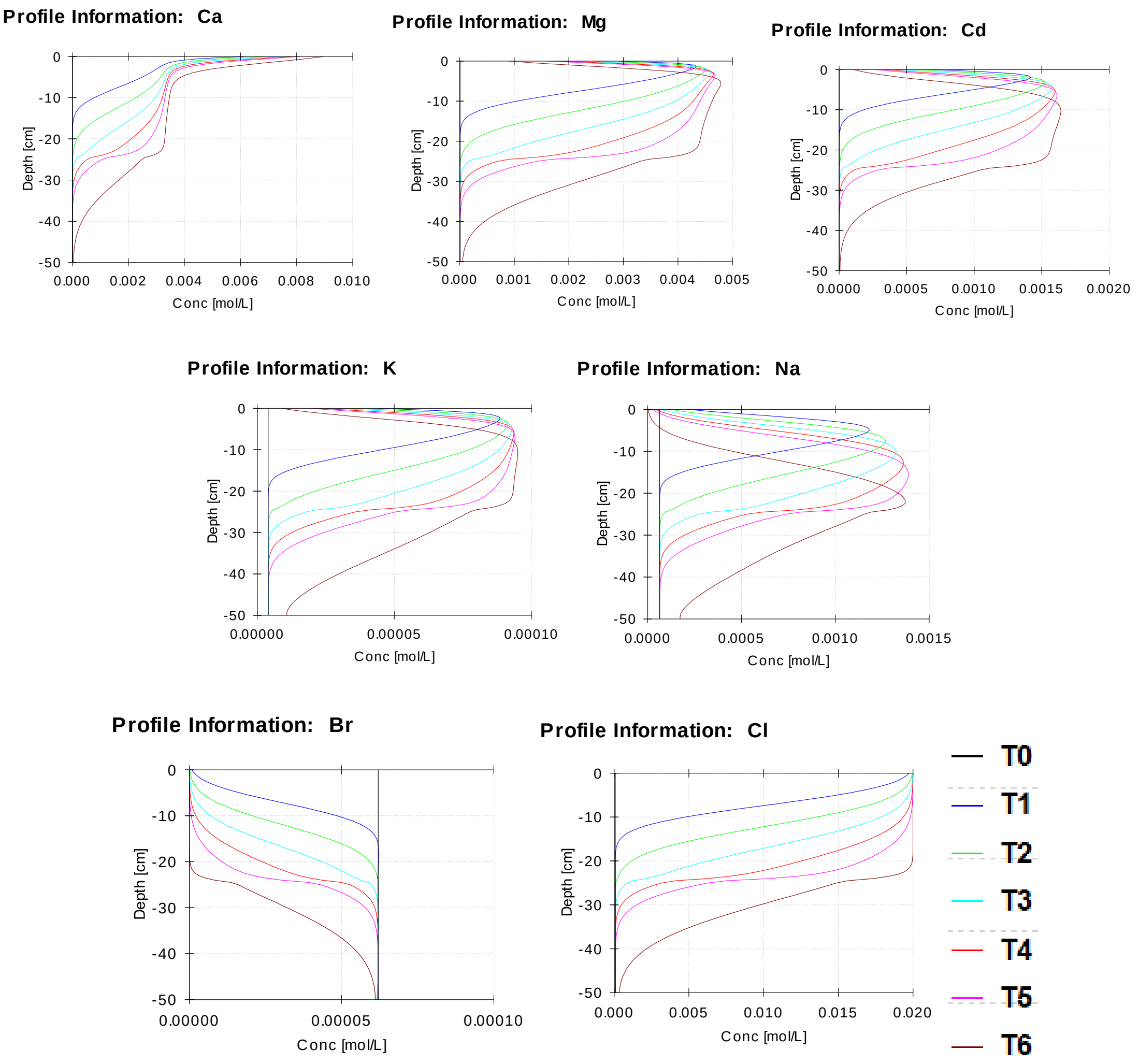

**Fig. 3 concentrations of the cation with +2 charge (up), cations with +1 charge (middle), and ions with -1 charge in different times and depths in the two-layered soil profile**

## 3.2 Two-layered soil with a tile drainage system

Results showed that water content and pressure head in the soil increased as the water flow through the system (Figures 4 and 5). Similar to the previous scenario, observation nodes below layer interface had higher water content since they comprised a finer soil. As can be seen from the fluxes chart, the fluxes of upper observation nodes were higher than below nodes. One reason for this result was that the subsurface drainage system collects the infiltrated water from the soil profile and drain water out of the soil column. Accordingly, the fluxes of the below nodes were influenced by the drainage system.

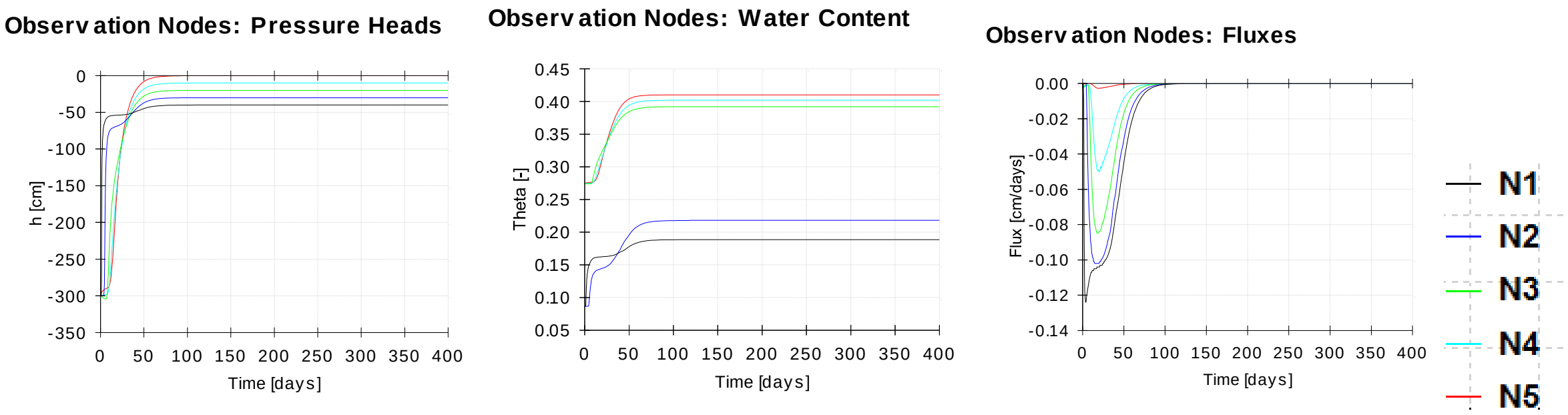

**Fig. 4 Seven variations of pressure heads and water content in different observational nodes of the two-layered soil with a tile drainage system**

The results of solute concentrations in the soil indicated that observation nodes below layer interface had the same concentrations in time and depth. The results of observation nodes below layer interface overlaid on observation node 3. This result arises from the fact that most of the solutes in the soil were drained out of the system by the drainage system. Results showed that solutes first accumulated between depth 10 and 20 cm and then washed out from the system by the drainage system. Therefore there was a decrease in solute concentrations during the study period.

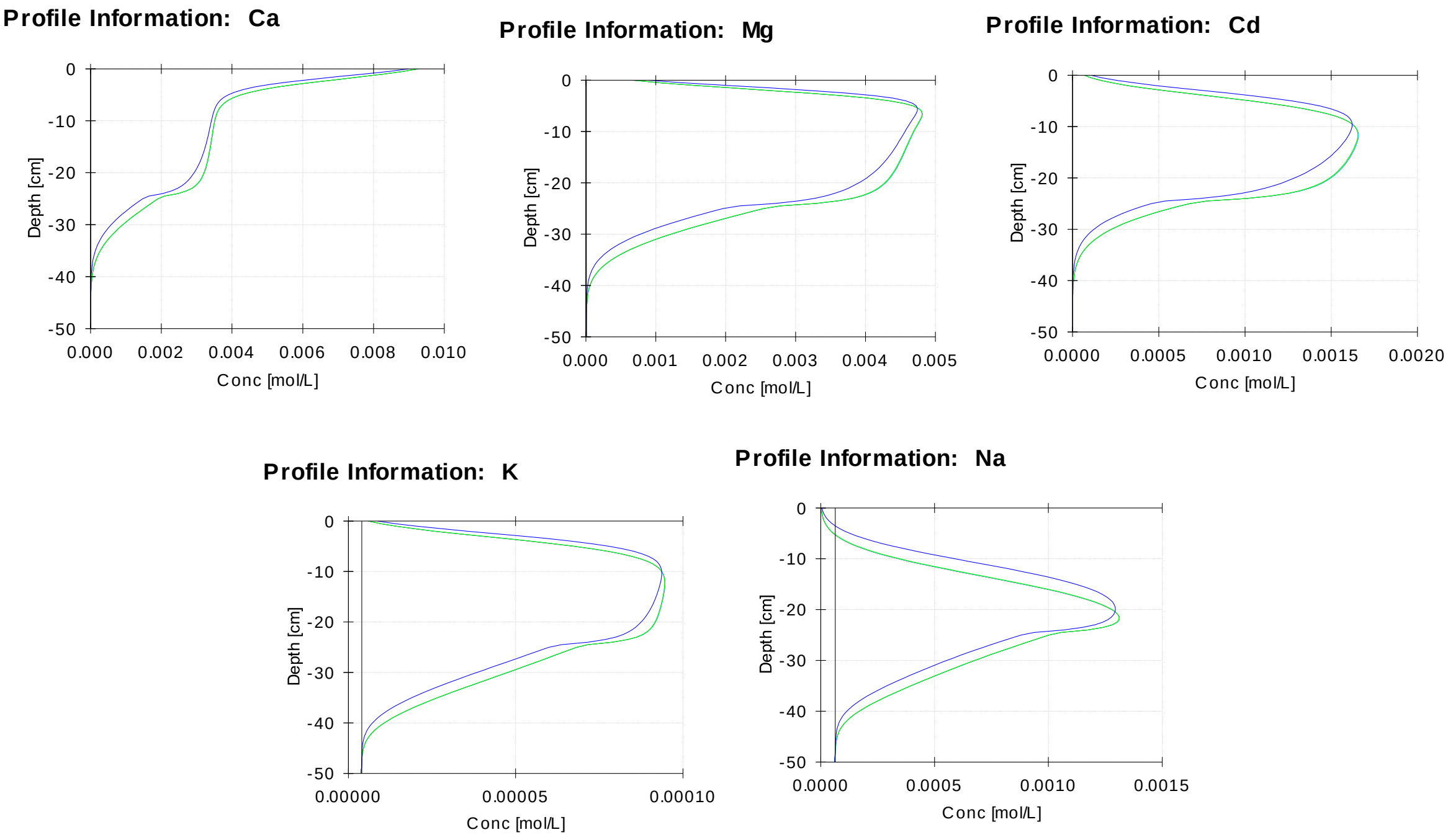

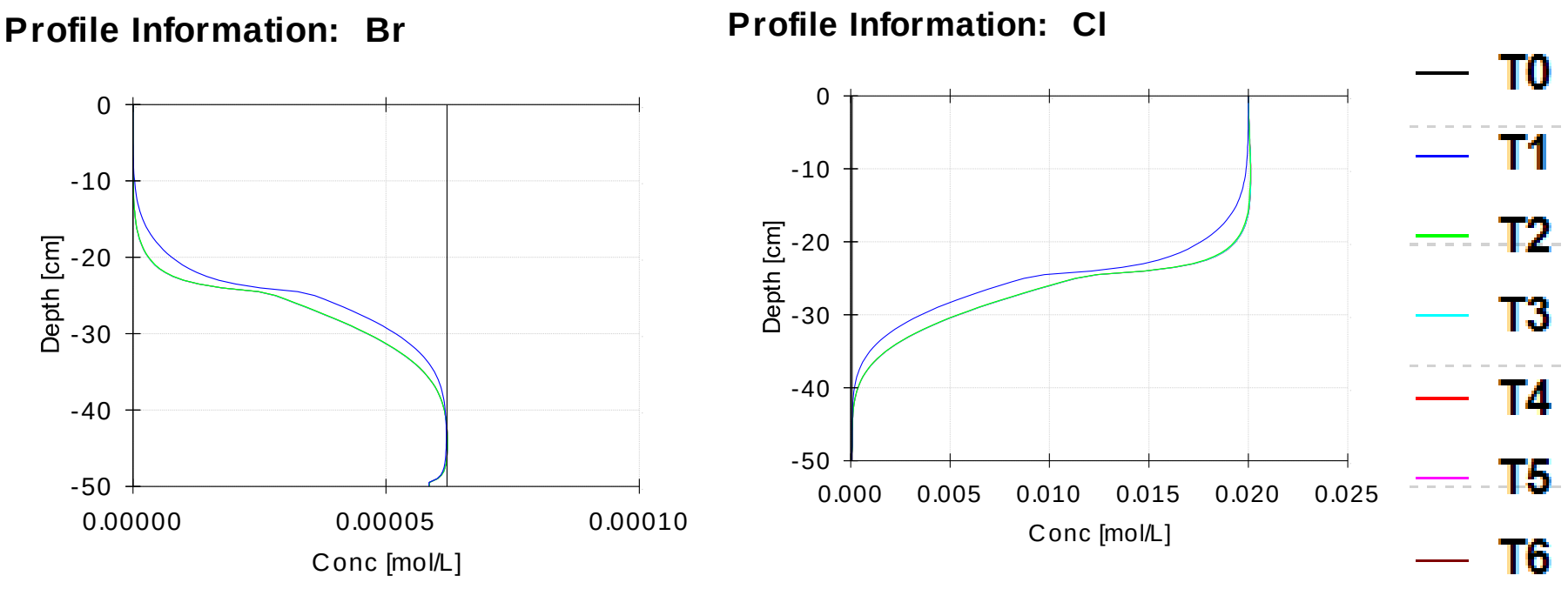

**Fig. 5 concentrations of the cation with +2 charge (up), cations with +1 charge (middle), and ions with -1 charge in different times and depths in the two-layered soil with a tile drainage system**

# 4 Conclusion

In this project, the effects of the presence of a soil layer and a subsurface drainage system on the simulation of water flow and solute transport using HP1 was presented. Results showed that soil layer change transport of chemicals by retarding their movement in the layer interface. This retardation resulted in an anaerobic condition, which accordingly causes a reduced environment. Additionally, the presence of a subsurface drainage system flushed out the excess amount of water and chemicals in the soil and thus decrease the water content and solute concentration in soil nodes below the subsurface drainage system. The results of this project showed that soil heterogeneities and bottom boundary conditions play an essential role in the prediction of chemical transport in the vadose zone.

# 5 References

Araghinejad S, Fayaz N, Hosseini-Moghari SM (2018) “Development of a Hybrid Data Driven Model for Hydrological Estimation”. Water Resources Management, 32(11), 3737-3750.

Hansen, D. J., McGuire, J. T., & Mohanty, B. P. (2011). Enhanced biogeochemical cycling and subsequent reduction of hydraulic conductivity associated with soil-layer interfaces in the vadose zone. Journal of environmental quality, 40(6), 1941-1954.

Jacques, D., & Šimůnek, J. (2005). User manual of the multicomponent variably-saturated flow and transport model HP1. Description, Verification and Examples, Version, 1, 79.

Jacques, D., Šimůnek, J., Mallants, D., & Van Genuchten, M. T. (2006). Operator-splitting errors in coupled reactive transport codes for transient variably saturated flow and contaminant transport in layered soil profiles. Journal of contaminant hydrology, 88(3), 197-218.

Jacques, D., Šimůnek, J., Mallants, D., & Van Genuchten, M. T. (2008a). Modeling coupled hydrologic and chemical processes: Long-term uranium transport following phosphorus fertilization. Vadose Zone Journal, 7(2), 698-711.

Jacques, D., Šimůnek, J., Mallants, D., & Van Genuchten, M. T. (2008b). Modelling coupled water flow, solute transport and geochemical reactions affecting heavy metal migration in a podzol soil. Geoderma, 145(3), 449-461.

Leterme, B., & Jacques, D. (2015). A reactive transport model for mercury fate in contaminated soil—sensitivity analysis. Environmental Science and Pollution Research, 22(21), 16830-16842.

Liu, X., Šimůnek, J., Li, L., & He, J. (2013). Identification of sulfate sources in groundwater using isotope analysis and modeling of flood irrigation with waters of different quality in the Jinghuiqu district of China. Environmental earth sciences, 69(5), 1589-1600.

Moayedi, H., Mosallanezhad, M., Rashid, A. S. A., Jusoh, W. A. W., & Muazu, M. A. (2018). A systematic review and meta-analysis of artificial neural network application in geotechnical engineering: theory and applications. Neural Computing and Applications, 1-24.

Mokari, E., Shukla, M. K., Šimůnek, J., & Fernandez, J. L. (2019). Numerical Modeling of Nitrate in a Flood-Irrigated Pecan Orchard. Soil Science Society of America Journal, 83(3), 555-564.

Schonsky, H., Peters, A., Lang, F., Abel, S., Mekiffer, B., & Wessolek, G. (2013). Sulfate transport and release in technogenic soil substrates: experiments and numerical modeling. Journal of Soils and Sediments, 13(3), 606-615.

Sedaghatdoost, A., Ebrahimian, H., & Liaghat, A. (2019). An Inverse Modeling Approach to Calibrate Parameters for a Drainage Model with Two Optimization Algorithms on Homogeneous/Heterogeneous Soil. Water resources management, 33(4), 1383-1395.

Sedaghatdoost, A., & Ebrahimian, H. (2015). Calibration of infiltration, roughness and longitudinal dispersivity coefficients in furrow fertigation using inverse modelling with a genetic algorithm. Biosystems Engineering, 136, 129-139.

Sedaghatdoost, A., Ebrahimian, H., & Liaghat, A. (2018). Estimating soil hydraulic and solute transport parameters in subsurface drainage systems using an inverse modelling approach. Irrigation and drainage, 67, 82-90.

Šimůnek, J., Jacques, D., van Genuchten, M. Th. & Mallants, D. 2006a. Multicomponent geochemical transport modeling using the HYDRUS computer software packages. J. Am. Water Resour. Assoc., 42(6), 1537-1547.

Zhang, H., Nordin, N. A., & Olson, M. S. (2013). Evaluating the effects of variable water chemistry on bacterial transport during infiltration. Journal of contaminant hydrology, 150, 54-64.